\documentclass[12pt,english]{apa6}
\usepackage{mathptmx,relsize}
\usepackage[utf8x]{inputenc}
\usepackage{natbib}

\makeatletter

\providecommand{\tabularnewline}{\\}

\makeatother

\usepackage{babel}
\begin{document}

\title{%
Availability of titles on peer-to-peer file sharing networks%
}
\shorttitle{%
Availability of titles on P2P file sharing networks%
}

\author{%
%---%
Petrus H Potgieter%
}

\affiliation{%
%---
Department of Decision Sciences, University of South Africa (Pretoria)\\
PO Box 392, Unisa, 0003, +27-12-429-4780\\
php@member.ams.org, potgiph@unisa.ac.za%
}

\abstract{%
File sharing, typically involving video or audio material in which
copyright may persist and using peer-to-peer (P2P) networks like BitTorrent,
has been reported to make up the bulk of Internet traffic \citep{pouwelse_pirates_2008,6011746}.
The free-riding problem appears in this ``digital gift economy'' but
its users exhibit rational behaviour \citep{becker_dynamics_2006}, 
subject to the characteristics of the particular network \citep{feldman_free-riding_2006}.
The high demand for the Internet as a delivery channel for entertainment
\citep{alleman_next_2009} underlines the importance of understanding
the dynamics of this market, especially when considering possible
business models for future pricing or licensing regimes \citep{gervais_price_2004}
and for the provisioning of network capacity to support future services.
The availability of specific titles on file sharing networks is the
focus of this paper, with a special emphasis on the P2P protocol BitTorrent. 
%Protocol design creates incentives that determine, among other things, the range of titles available on a given platform.
The paper compares
the incentives provided in BitTorrent to those in other file-sharing
communities, including file hosting, and discusses the number of titles
available in the community at any given time, with an emphasis on
popular video items with ambiguous legal status \citep{Watters201179}.
}

\maketitle

\section{Introduction}

The main object of this research is to understand how incentives operate
in the world of (mainly anonymous) file sharing, to produce a spectrum
of available content. The mechanism for sharing will clearly influence
the amount of content available to any specific individual seeking
such content online. After all, the proliferation of the technology
for home audio and video taping in the 1970s and 1980s created an
environment in which any individual had easy access to all material
under the control of their conventional social peers but not more.
The legal controversy \citep{lessig_freeing_2004} of then and now
is however not the main topic of this work.

The Internet, together with digital multimedia formats, has enabled
media sharing (like many other things) to take place between complete
strangers all over the globe and with lossless transmission of the
content. The use of file-sharing networks for distributing unauthorised
copies of copyrighted material is of concern to the media and publications
industries \citep{doi:10.1080/01972240802189468} and one reaction
has been attempts by authorities in many countries to either block
file-sharing traffic or to ban access to websites indexing such content,
recently in Italy and reportedly even more stringently in the United
Kingdom \citep{arthur_internet_2010}. Given the high percentage of
Internet traffic involving file sharing \citep{gummadi_measurement_2003,menasche_content_2009,plissonneau_analysis_2005,pouwelse_pirates_2008}
this is of particular concern also in the debate around network neutrality.

‭The high demand for the Internet as a delivery channel for audio-visual
entertainment‭ \citep{alleman_next_2009} underlines the importance
of understanding the dynamics of this market,‭ especially when considering
possible business models for future pricing or licensing regimes‭
\citep{gervais_price_2004} and for the provisioning of network capacity
to support future services.‭ This is of particular interest when it
becomes a matter of public policy, for instance where tax-funded investment
in new generation networks (NGNs) are being considered. 

The availability of specific titles on file sharing networks is the
focus of this paper,‭ with a special emphasis on the peer-to-peer
(P2P) protocol BitTorrent. An overview of the operation of the network
will be given and we shall consider the model proposed by \citet{menasche_content_2009}
for content avaibility in BitTorrent swarms. Alternatives to P2P are
briefly considered and finally some factors determining the availability
of titles are discussed, together with data illustrating users' behaviour
on the network.

\section{Altruism in file-sharing}

File sharing communities on the Internet are characterised by a high
degree of anonymity, especially in light of the (somewhat remote)
possibility of prosecution or (more likely) threatening letters from
Internet service providers (ISPs). Indeed, these communities are responding
to attacks from copyright holders by evolving more sophisticated mechanisms
for maintaining anonymity. Unlike the taping and mix-taping of earlier
decades, where the sharing communities presumably coincided with ordinary
social networks, file sharing requires a degree of altruism that is
not backed up by an implicit offline social quid-pro-quo convention. Nevertheless,
a lot of file sharing between strangers evidently does take place.
The rôle of altruism in these networks has been studied by many people,
e.g. \citet{feldman_free-riding_2006}. Basically, small costs can
be imposed for free-riding. Furthermore, there are closed networks
where this is not a problem and users exhibit rational behaviour within
this‭ ‬digital gift economy \citep{becker_dynamics_2006} subject
to the characteristics of the particular network‭ \citep{feldman_free-riding_2006}‬.‭ 

The principle of P2P networking is illustrated by the following toy
example. Suppose that a single publisher%
\footnote{Here \emph{publisher} signifies any entity or individual making a
file available to others, and not necessarily the copyright owner
or offline publisher.%
} of some specific content appears on the network where a single potential
recipient is waiting. If the recipient (or, \emph{leecher}) is not
prepared to donate anything to others on the network, and the publisher
(or, \emph{seeder}) is prepared to donate only one copy, the net effect
will be that the leecher downloads a single copy and the publisher
donates one copy. Further leechers arriving on the network and seeking
a copy will not be serviced. However, should leechers be prepared
to act as peers, i.e. to let other download from them, then a large
number of copies can be distributed from a single seeder. Suppose
$n$ leechers appear simultaneously, while the original seeder is
available, and each leecher is prepared to donate only the equivalent
of $\frac{n-1}{n}$ copies of the file. Then, it becomes possible
for each of the leechers to obtain the entire file in the following
way.
\begin{enumerate}
\item Let the $k$-th leecher download the $k$-th part of the file, of
size $\frac{1}{n}$, so that the seeder will have donated only a total
of one full copy of the file.
\item Now, let each leecher donate his/her fraction $\frac{1}{n}$ of the
desired file to each of the $n-1$ other leechers.
\end{enumerate}
At this point each leecher will have obtained a full copy of the file
but will have uploaded less than a full copy of the content.

It is easy to see how some free-riding can be accommodated within
the system. Suppose there were, as above, $n$ peers and one seed
but also a single leecher not willing or able to upload any content.
If each of the $n$ original peers is prepared to donate a full copy
of the content, only step 2 above need be modified to enable the free-riding
leecher to also obtain a full copy of the content. Obviously many
more free-riding leechers could be serviced, should sufficiently many
peers be prepared to make available even more than a single multiple
of the content. Equally obviously, the ability of leechers to download
desired content is entirely dependent on the willingness of peers
to make the content available to others.

The toy example is simplistic and does not even incorporate the influence
of accidental sharing by leechers. In the case of very popular content,
most P2P clients will automatically share chunks of content already
downloaded while not yet in possession of the complete file and will
continue seeding the content once the download has been completed
until either
\begin{itemize}
\item a prescribed share ratio (number of full copy equivalents shared to
others) has been reached, or
\item the client or torrent is stopped manually by the user.
\end{itemize}
In the case of popular content just after publication, many leechers
will unintentionally act in a very altruistic way since they will,
while waiting for a download to complete, facilitate many uploads
of the same material to others, also since the demand is high at these
times. Such unintentional peering probably contributes significantly
to the availability of titles shortly after release, for example popular
television shows in the days after which they have been broadcast
or otherwise leaked to the public. Even though the preceding illustrates
that free-riding can be accommodated in a file sharing community,
the presence of a community of BitTorrent publishers with a financial
incentive has been hypothesized \citep{Cuevas:2010:CPB:1921168.1921183}.

Consider the (perhaps rather unscientific) statistics published by
the website KickAssTorrents.com in Table \ref{tab:Availability-on-PublicBT.com},
harvested from the BitTorrent tracker PublicBT.com for episodes from
Season 6 of the series Desperate Housewives published through the
highly regard and prolific EZTV. Availability of Episode 1 seems good
even after several months although the amount of leeching going on
at a specific time is not very high at all, the number of downloaders
being just over 1\% of the number observed less than 48 hours after
episode 20 of the show was shown on ABC broadcast television in the
United States. Clearly the amount of free-riding or miserly sharing
is much higher early in the life of the swarm.%
\footnote{A single title can be available in many swarms \citep{VinkoEA}. The
statistics in Table \ref{tab:Availability-on-PublicBT.com} are for
a single swarm.%
}

\begin{table}
\caption{\label{tab:Availability-on-PublicBT.com}Availability on PublicBT.com
for two selected episodes of Desperate Housewives}
\small
\begin{centering}
\begin{tabular}{|llrr|}
\hline
File name & Date %and time 
& Seeders & Leechers\tabularnewline[\doublerulesep]
\hline 
\ldots{}S06E20\ldots{} & 27 April 2010%, 15:32:30 
& 10\,113 & 3\,863\tabularnewline
\ldots{}S06E01\ldots{} & 27 April 2010%, 15:32:24 
& 334 & 48\tabularnewline
\hline 
\end{tabular}
\end{centering}
\end{table}

\section{The BitTorrent network}

The BitTorrent protocol%
\footnote{Invented by Bram Cohen.%
} enjoys widespread use and has been implemented on many platforms.
It is also the primary protocol supported by the famous PirateBay
portal for online content and subject of frequent legal action.‭ However,
BitTorrent (BT) is also used for distributing software such as installation
discs for new Linux distributions and for scientific data \citep{langille_biotorrents:file_2010}.
The main reason for the success of BT is that it scales very well
when demand increases. As outlined cursorily above, a P2P distribution
system can easily accommodate a very high demand for a certain large
file of collection of files without a specifically high degree of
investment at any specific node \citep{izal_dissecting_2004}. In
fact, free-riding appears to be the only problem other than unavailability
of content. It will be helpful for the discussion below to describe
briefly the operation of a BT network, from the user's point of view.

A prospective user of general content will typically visit a torrent
index site such as PirateBay%
\footnote{http://thepiratebay.org/%
} or AnimeSuki%
\footnote{http://www.animesuki.com/%
} where the potential downloader will click on a file with the .torrent
file extension (or, more recently, a .magnet file). This torrent file
will contain references to specific torrent tracking servers and might
well be opened automatically by BT client software%
\footnote{For example, Vuze or BitTornado.%
} on the user's computer so that the user will not necessarily even
be aware of the torrent file per se. The BT will start querying the
trackers listed in the torrent file as part of the process of joining
a swarm. The tracker servers will provide information about actual
peers already in the swarm and the new peer will start requesting,
and eventually offerings, parts of the file to be download. The download
time experienced by the user depends very much on the conditions in
the swarm \citep{chiu_minimizing_2008}.

\subsection{The BT swarm as $M/G/\infty$ queue}

\citet{menasche_content_2009} describe a BT swarm as an $M/G/\infty$
queue \citep{browne_transient_1993}. That is, the swarm is supposed
to consist of publishers which appear at irregular intervals and peers
with Poisson arrivals. The queuing system is busy as long as the title
remains available from peers that are online. They deduce estimates
for the expected length of a busy period under various assumptions
w.r.t. peer behaviour. For example, if all peers are selfish and leave
as soon as completing their download has been completed, the expected
length of a busy period is
\[
B=\frac{e^{s(r+\lambda)/\mu}-1}{r+\lambda}
\]
where $s$ is the size of the file, $\mu$ is the mean download rate
of peers, $r$ the arrival rate of publishers and $\lambda$ the arrival
rate of peers. It is further assumed that peers who arrive when the
content is not available, immediately leave. Publishers are assumed
to stay online only for a time $\frac{s}{\mu}$, i.e. long enough
to serve one copy of the content. The reader will be able to confirm,
at a glance, that $B$ behaves as common sense would suggest when
$\mu\rightarrow\infty$ and $\lambda\rightarrow\infty$. One can also
observe that doubling the size of the file from $s$ to $2s$ would
increase the size of $B$ by a factor of
\[
\frac{e^{2s(r+\lambda)/\mu}-1}{e^{s(r+\lambda)/\mu}-1}
\]
which is very substantial. The busy periods will generally be much
longer in a swarm with file size $2s$ than the sum of the expected
busy times of two separate swarms with file size $s$. \citet{menasche_content_2009}
obtain much more precise mathematical estimates of the advantage in
terms of availability in large swarms and could thereby present a
convincing explanation of content bundling%
\footnote{Content bundling is the phenomenon where individual books or episodes
of a television series are not available as single files but only
in a bundled torrent consisting, for example, of hundreds of e-books
on a particular topic, or of all of a specific season of a television
series.%
}.

Predictions by the model of \citet{menasche_content_2009} are consistent
with a vast amount of data on real networks investigated in their
study. The data in their work showed 80\% of swarms as \emph{unavailable}
at least 80\% of the time. Nevertheless, their model might be inappropriate
for the large amount of BT activity that consists of the distribution
of material within the first few weeks (or days) after it appears.
If this initial burst of activity is take into account, the arrival
rate of publishers and peers is certainly far from constant, as one
can clearly see in Table \ref{tab:Torrent-activity-for} where it
seems clear that activity peaked about 36 hours after the swarm was
constituted and the declined rather rapidly. The $M/G/\infty$ queue
also has no room for incorporating individual behaviour or system-wide
constraints such as the total amount of available storage and bandwidth
and possibly limited demand for audio-visual content.

\begin{table}
\begin{centering}
\begin{tabular}{lrr}
Date and time & Seeding & Leeching\tabularnewline
\hline 
27 April 2010, 15:32:30 & 10\,113 & 3\,863\tabularnewline
27 April 2010, 19:15:01  & 11\,187 & 5\,132\tabularnewline
27 April 2010, 21:19:10 & 11\,000 & 4\,872\tabularnewline
27 April 2010, 22:31:42 & 9\,664 & 4\,468\tabularnewline
28 April 2010, 22:16:10 & 7\,701 & 2\,445\tabularnewline
29 April 2010, 05:20:27 & 4\,640 & 1\,078\tabularnewline
29 April 2010, 21:32:32 & 6\,825 & 1\,887\tabularnewline
30 April 2010, 07:35:36 & 4\,416 & 840\tabularnewline
\end{tabular}
\par\end{centering}

\caption{\label{tab:Torrent-activity-for}Torrent activity for Desperate Housewives
S06E20 during the week after its first broadcast, from kickasstorrents.com,
reported for tracker.publicbt.com and a swarm formed on 26 April 2010
with purposive sampling.}

\end{table}

\subsection{A fluid model}

\citet{qiu_modeling_2004} use partical differential equations to
describe a fluid model for BT swarms. As om the model of \citet{menasche_content_2009}
the parameters like peer arrival are assumed to be constant. They
prove the existence of a Nash equilibrium under certain conditions
and have experimental data that appears to validate their model under
the given assumptions.

\section{Anonymous file hosting}

Anonymous file hosting is another way of sharing content when the
publisher does not necessarily want to do so openly, possibly out
of fear of political persecution, harrassment or of prosecution for
possibly copyright violations. An anonymous file host allows users
to upload a file to an Internet web page with a generic name%
\footnote{http://rapidshare.com/files/16433818/ for example.%
} that gives no indication of the content of the file stored there.
The uploader might place a link to the web page in an Internet forum
or circulate it in another way. Registration is not required of casual
users but the business model of these providers of \emph{one-click
hosting} evidently includes enticing users to take out a subscription
which allows downloading files without the waiting time imposed on
casual users of the free service. Subscribers also enjoy faster download
speeds and file hosting offers a far greater degree of anonymity than
P2P distribution \citep{le_blond_angling_2010,le_blond_spyingworldyour_2010}
i.a. since both publishers and peers are exposed for only as long
as it takes to transfer the file from/to the hosting site. It is quite
clear that one-click hosting has created a revenue model for content
sharing, whether legal or possibly illegal, and \citet{antoniades_one-click_2009}
observe that of a list of 100 unpopular film titles, more are available
on RapidShare than on BitTorrent.

\section{Title availability}

In this section, we consider the characteristics of an ideal and fairly
complete model of title availability. First, consider the salient
features of the networked digital multi-media world.
\begin{description}
\item [{Hollywood\ universality}] \citet{antoniades_one-click_2009} found
100\% availability on PirateBay for the top 50 US DVD rental titles
for the week under investigation but only 76\% availability for Amazon's
top 25 German films of all time. Anecdotal evidence suggests that
popular US television series become available within a few hours of
airing and are nearly universally available because of the ease of
digital home recording.
\item [{Ease\ of\ copy}] The cost of copying a computer file is nearly
zero and does not degrade the original copy. Hence, even compared
to relatively inexpensive media such as CD or DVD discs, the marginal
cost of reproduction is exceptionally low for computer copies. Further,
digital rights management (DRM) is not particularly efficient%
\footnote{Not least of all because of the \emph{analog hole.}%
} and there is no limit to the number of copies of a particular item
a consumer can make. As more consumers source their entertainment
from unencrypted digital sources, the smaller the impact of DRM will
become.
\item [{Storage~efficiency}] Digital media can be quite efficiently stored.
A single portable computer hard disk can store hundreds or thousands
of films in a space no greater than that of a single DVD box.
\item [{Open~borders}] There are no customs inspections on the Internet
and with relatively inexpensive bandwidth in many places, the cost
of transmitting digital content is low. Storage capacity of offline
media has not increased sufficiently quickly in order to make the
physical transport of media files an efficient proposition.
\end{description}
In view of the German film example, we suppose only that almost all
English-language content produced for US television or cinema audiences,
is available in digital format on some computer which is connected
to the Internet. Such material would, for many, constitute a perfectly
acceptable source of entertainment, perhaps to the extent that they
would need no other. It is not quite the case yet but one might also
assume that each potential consumer has sufficient storage space available
to archive a full copy of this corpus. The corpus grows only slowly
and perhaps slowly enough that a typical residential user in the US
could download the entire accrual each day.

Assume, for argument's sake, that Hollywood is somehow wiped out and
that the corpus stops growing. In this case, it would be of benefit
to each user to download the entire corpus and this would be possible,
in principle, using P2P if each user is only prepared to contribute
approximately as much as they download. This would be a real socialist
utopia, but for the problem of enforcing cooperation and punishing
free-riding. Since even BT is subject to opportunistic strategic manipulation
\citep{levin_bittorrent_2008} the availability of titles is still
very much subject to constraints on storage, network use and the problem
of free-riding. However, many tests have shown BT to be an efficient
allocator of resources and P2P could very well be one of the mechanisms
that would assist in a fair allocation of network resources on the
Interent, a network not initially designed for commercial use.

In the discussion above, it has been taken for granted that consumers of content have a fixed preference in terms of quality. That is, given a specific unit of artistic content (an episode of Desperate Housewives, for example) corresponds to a specific chunk of data with associated storage and transmission costs. However, the proliferation of high-definition display devices means that consumers now increasingly prefer high-definition content. High-definition video requires file sizes that are several times those of standard definition content and anecdotal evidence suggest that the spectrum of content available in high-definition formats (1080p or 720p) is still far below that of the old standard television resolution.

\section{Conclusion}

Protocol design creates incentives that determine,‭ among other things,‭ the range of titles available on a given platform.‭ Online sharing involves relatively high costs (e.g. the decrypting, copying and recoding or “ripping” of a commercial DVD in a compressed and unencrypted digital format) incurred by publishers of material and much smaller costs
for those who simply share files which they have already downloaded. The world of online file-sharing is not simply a free-for-all of illegal material but rather an almost organic self-organising economy which manages to allocate scarce resources and supply consumer demand. It has even been suggested that digital downloads often help to increase
sales offline \citep{peitz_music_2006}. The modelling of P2P networking presents a fascinating opportunity for network economists and applied mathematicians, with much to be done to formulate a realistic model that properly incorporates the dynamics of the system. Observing the spread of high-definition content will provide further data but will also require a model of consumers' preference for high-quality content.

%\raggedright

\bibliographystyle{apalike}
\bibliography{PHP_P2P}

\end{document}